\documentclass[10pt,a4paper,twocolumn]{article}
\usepackage[english]{babel} 
\usepackage[top=2.8cm,bottom=2.8cm,left=2.cm,right=2.cm]{geometry}  
\usepackage{titlesec}
\usepackage{color,soul}
\usepackage[utf8]{inputenc} 
\usepackage[affil-it]{authblk}
\usepackage{amsmath,amsfonts,amssymb}
\usepackage{varioref}
\usepackage[colorinlistoftodos]{todonotes}
\usepackage[pdftex]{hyperref} 
\let\OLDthebibliography\thebibliography
\renewcommand\thebibliography[1]{
  \OLDthebibliography{#1}
  \setlength{\parskip}{1pt}
  \setlength{\itemsep}{1pt plus 0.3ex}
}
\usepackage[binary-units = true]{siunitx}
\usepackage{float}
\usepackage{tikz}
\usepackage{xcolor}
\usepackage{comment}
\usepackage{braket}
\usepackage{nicefrac}
\usepackage{caption}
\usepackage{graphicx}  
\graphicspath{{./Figures/}{./PSTricks/}}  
\usepackage{multirow, makecell}
\usepackage{appendix}
\usepackage{setspace}
\usepackage{cite}
\usepackage{csquotes}
\usepackage{array,booktabs}

\usepackage{enumitem}
\usepackage{caption3} 
\DeclareCaptionOption{parskip}[]{}
\usepackage[font=small,labelfont=bf]{caption}
\usepackage[final]{pdfpages}
\usepackage[colorinlistoftodos]{todonotes}
\usepackage{tikz}
\usepackage{abstract}

\usepackage[export]{adjustbox}
\usepackage{dblfloatfix}
\usepackage{comment}

\begin{document}
%\maketitle
\title{\Large \textbf{A proposal for practical multidimensional quantum networks}}
\author[1,*]{\small Davide Bacco}
\author[2]{\small Jacob F. F. Bulmer}
\author[3,4]{\small Manuel Erhard}
\author[4,5,6]{\small Marcus Huber}
\author[7]{\small Stefano Paesani}
\affil[1]{\footnotesize CoE SPOC, DTU Fotonik, Technical University of Denmark, 2800 Kgs. Lyngby, DK}
\affil[2]{\footnotesize Quantum Engineering Technology Labs, University of Bristol, Bristol BS8 1FD, UK}
\affil[3]{\footnotesize Vienna Center for Quantum Science \& Technology (VCQ),Faculty of Physics, University of Vienna, AT.
}
\affil[4]{\footnotesize Quantum Technology Laboratories (qtlabs) GmbH, Wohllebengasse 4/4 1040 Vienna, AT}
\affil[5]{\footnotesize Institute for Atomic and Subatomic Physics, Vienna University of Technology, Vienna, AT
}
\affil[6]{\footnotesize Vienna Center for Quantum Science and Technology, Atominstitut, TU Wien, 1020 Vienna, Austria}
\affil[7]{\footnotesize Center for Hybrid Quantum Networks (Hy-Q), Niels Bohr Institute, University of Copenhagen, Blegdamsvej, DK}

\affil[*]{dabac@fotonik.dtu.dk}

\date{} 
\pagestyle{plain}
\setcounter{page}{1}
\twocolumn[\begin{@twocolumnfalse}
\maketitle
     \vspace{-0.8cm}
\begin{abstract}
\normalsize
\vspace*{-1.0em}
\noindent
A Quantum Internet, 
i.e., a global interconnection of quantum devices, 
is the long term goal of quantum communications, and has so far been based on two-dimensional systems (qubits).
Recent years 
have seen 
a significant development 
of high-dimensional quantum systems (qudits).
While qudits present 
higher photon information efficiency 
and robustness to noise, their use in quantum networks 
present experimental challenges 
due to the impractical resources 
required in high-dimensional quantum repeaters.
Here, we show that such challenges 
can be met via the use of standard quantum optical resources, such as weak coherent states or weak squeezed states, and linear optics.
We report a concrete design and simulations of an entanglement swapping scheme for three and four dimensional systems, showing how the network parameters can be tuned to optimize secret key rates and analysing the enhanced noise robustness at different dimensions. 
Our work significantly simplifies the implementation of high-dimensional quantum networks, fostering their development with current technology.
\vspace{0.5cm}
\end{abstract}
\end{@twocolumnfalse}]

\section*{Introduction}
\vspace{-0.25cm}
The advent of quantum information has strongly influenced modern technological progress. Intense research activities have been carried out in the last two decades on such field, producing outstanding results, e.g. in quantum computing~\cite{preskill2018quantum}, communication\cite{Pirandola2019,Boaron2018_421km,Minder2019,bacco2019,Yin2016} and simulation\cite{paesani2019,georgescu2014} with the final goal of realizing a Quantum computer and a Quantum Internet. The Quantum Computer will for example enable accurate simulations of chemical and biological compounds, while the Quantum Internet will allow the communication between users (either classical or quantum) guaranteeing multiple applications, from secure communications to remote quantum computing\cite{divincenzo1997,wehner2018}. 
\begin{figure*}[th!]
\centering
\includegraphics[width=0.99\textwidth]{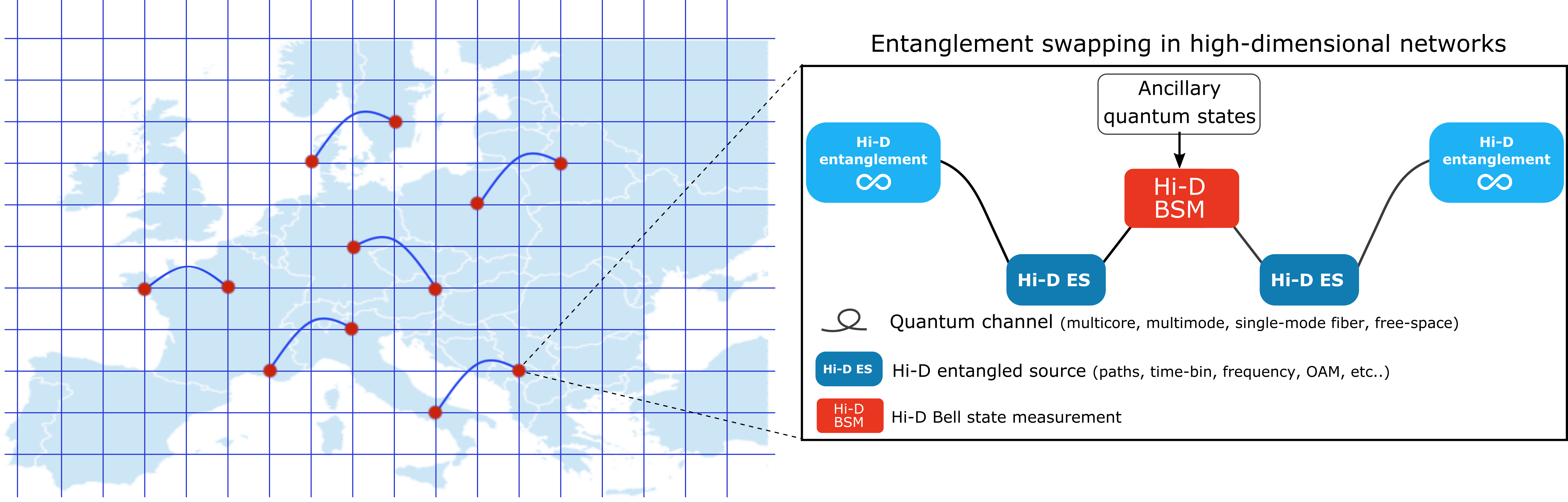}
\caption{{\bf Imaginative view of a multidimensional quantum network.} The inset describes a potential building block of a such a network. Two sources of entangled photons (cyan boxes) emit high-dimensional quantum states which are propagated through different quantum channels. An intermediate node (red box) measures the two multidimensional photons using ancillary quantum states. Once the Hi-D quantum interference generates a positive outcome, multidimensional entanglement is generated between the two users (light blue boxes).}
\label{fig:high_dimensional_network}
\end{figure*}

Both in the quantum computer infrastructure and in the quantum internet, it is crucial to transport quantum states- i.e. perform quantum communications-, either between components in the same quantum device\cite{Zhong2021} or between nodes in a network ~\cite{northup2014}. Independently from the technology which will master the challenge of realizing the quantum computer, photons are the only candidates for transmitting quantum information over long distances~\cite{Boaron2018_421km,Minder2019}.  In the case of continental applications, these photons are easily transportable by optical fibres, and since this medium is already deployed worldwide in the context classical communication, the quantum community plans to reuse the same fibre infrastructure. Unfortunately, the transmission of photons, over long distances, is limited by the intrinsic loss of the optical fibres, and by the external noise due to the interaction with the outside environment, which destroys the quantum states. Another hurdle comes from central laws of physics, which state the infeasibility to create an identical copy of an arbitrary unknown quantum state~\cite{wootters1982}. A solution to these problems is provided by Quantum Repeaters (QR), i.e., the quantum counterpart of classical repeaters~\cite{briegel1998, munro2012,Muralidharan2014, muralidharan2016}.  Different kinds of quantum repeater protocols can be identified and are usually classified into three distinct categories or as generations ($1^{st}$ generation: quantum memory and purification protocol, $2^{nd}$ quantum repeater with error correction against operational errors, $3^{rd}$ fully fault-tolerant quantum repeaters ~\cite{zukowski1993,munro2012,Muralidharan2014}). 
Current quantum repeater proposals mainly rely on two-dimensional encoding schemes (\textit{qubit}) as an information unit, which due to the decoherence processes, caused by interaction with the external environment, lose their ability to stay in superposition and/or in an entangled state. This can be directly translated to a limited robustness to noise and thus limit the overall applicability.

%Here, we report a theoretical proposal for a quantum network based on multidimensional quantum states, analysing pro and cons of such schemes. Furthermore, we report a concrete design and simulations of an entanglement swapping scheme, for a three and four dimensional systems, using state-of-art technique. Our simulations prove the advantage of multidimensional quantum states in the high-noise regime.

%\textcolor{blue}{New part starts here}
A potential workaround is provided by high-dimensional quantum states (qudits), which offer an intrinsic advantage in terms of photon information efficiency and robustness to noise~\cite{cozzolino2019review,ecker2019overcoming}.
In recent years many advances towards the development of high-dimensional (HiD) quantum networks have been achieved. High-quality generation and manipulation of high-dimensional entangled states have been demonstrated using the OAM, time-bin, frequency, and path degrees of freedom using bulk or fiber optics, as well as in integrated quantum photonics~\cite{erhard2020advances}. Low-loss transmission of high-dimensional states has been reported through fiber and free-space with high fidelities~\cite{cozzolino2019orbital}.

\begin{figure*}[h!]
\centering
\includegraphics[width=0.99\textwidth]{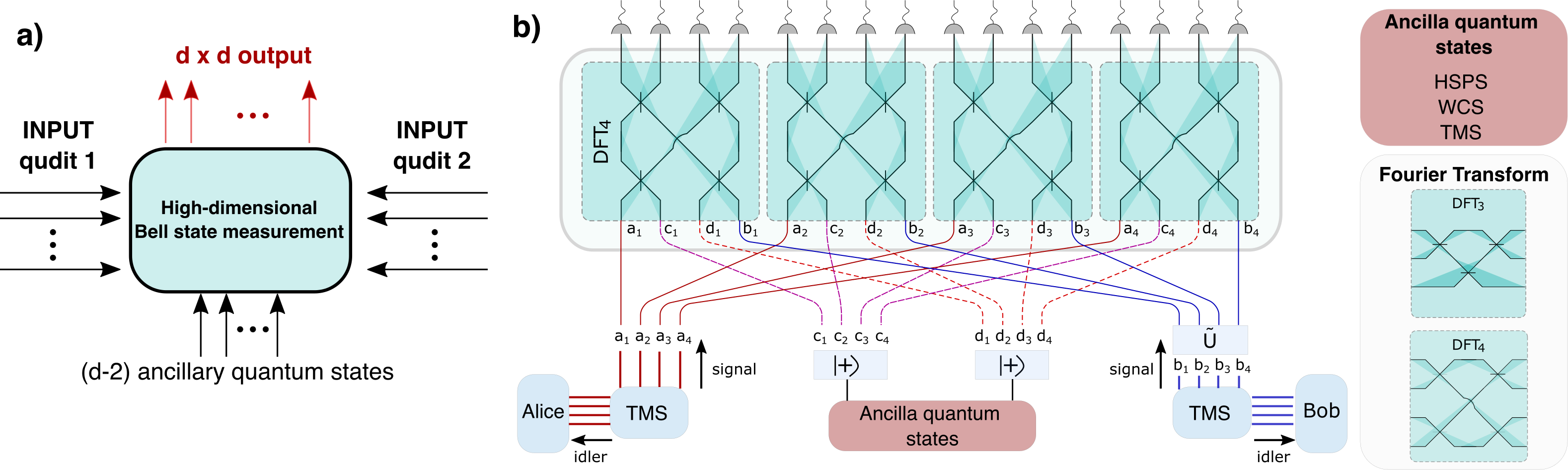}
\caption{{\bf Scheme of the high-dimensional quantum interference with linear optics.} {\bf a) Overview scheme.} Input 1 and input 2 are the two high-dimensional quantum states encoded in path degree of freedom ($d$ dimensionality of the Hilbert space). The scheme requires $d-2$ ancillary quantum states and $d^2$ outputs for a complete measurement. {\bf b) Overview scheme.} Detailed setup for the high-dimensional BSM. TMS: two-modes squeezed state, WCS: weak coherent state, HSPS: heralded single-photon source; DFT$_n$ discrete fourier transform in a n-dimensional system.}
\label{fig:bs}
\end{figure*}

However,the realisation of high-dimensional quantum repeaters has remained an open experimental challenge mainly due to the lack of schemes for Bell measurements in higher dimensions, i.e. a fundamental action in the quantum repeater schemes. 
Indeed, it has been proved that, without ancillary photons, the projection into a high-dimensional Bell state is unattainable with only linear optics~\cite{Calsamiglia2002}. %A possible route is the use of non-linear effects, which however remain highly inefficient with current technologies. 
Very recently, two different schemes have been proposed to perform linear-optical Bell measurements in high dimensions with the use of ancillary single photons, and demonstrated for $d=3$\cite{luo2019quantum,hu2020experimental}.  However, the lack of deterministic single photon sources makes the requirements of such ancillary single photons a stringent hurdle for current technologies, limiting the practicability of developing high-dimensional quantum networks.

Here, we show that schemes for high-dimensional Bell measurements can operate with high-fidelity, using practical resources: weak-coherent states and sources of weak squeezing. We numerically test the performances of this scheme for realistic quantum repeaters by developing a simulator of the Gaussian resources and evolutions employed, and single-photon measurements with the use of threshold detectors.
Even though the experimental resources are significantly simplified, we numerically identify regimes where high-dimensional quantum states offer advantages in entanglement-based quantum networks protocols. 
Our proposal paves the way for a practical implementation of high-dimensional quantum states in future quantum networks.

\section*{Protocols}
\label{sec:protocols}
Future quantum networks will be constituted of quantum nodes (i.e. repeater station) where non classical correlation will be shared between different users. In order to achieve this goal, it is possible to define three main operations, which are reported in Figure~\ref{fig:high_dimensional_network}: generation, transmission and interference of high-dimensional quantum states.
Regarding the generation and transmission of HiD quantum states, an increasing number of demonstrations have been achieved during the last five years. However, HiD Bell state measurements (BSM) remain still challenging due to its limited practicability and efficiency. %The use of non-linear optics will result in a very poor performance compared to the bi-dimensional case.

\paragraph*{Linear-optical circuit for Bell measurements in high-dimension.}
Let us consider the simplest building block for a quantum network based on high-dimensional quantum states, as depicted in Figure~\ref{fig:high_dimensional_network}. The goal is to extend the achievable distance by connecting two high-dimensional entangled quantum states with a repeater. Here we envision the simplest quantum repeater, an optical Bell measurement.

An optical Bell-measurement is a detectable projection of two single photons into a maximally entangled Bell state. For two-dimensional photons, the Bell measurement can be performed by a single beam-splitter and two
single-photon detectors at the outputs. The trick is to utilize two-photon interference at a 50/50 beam-splitter, namely Hong-Ou-Mandel interference,
and detect the two photons simultaneously after the beam-splitter in two different outputs. If such an event occurs, we know that the two photons have been projected into an asymmetric Bell state at the beam-splitter. This Bell-measurement is the key part of every photonic quantum teleportation or entanglement-swapping experiment as it can entangle two photons that never interacted before.

Luo et al.\cite{luo2019quantum} extended these ideas and proposed a generalized BSM in higher-dimensions. They replaced the beam-splitter with a generalized Fourier-transform interferometer (FTI). The FTI interferes different photons $\{a, b, c, d\}$ that all occupy the same mode $m$, e.g. $|1\rangle_a|1\rangle_c|1\rangle_d|1\rangle_b$ as depicted in Figure~\ref{fig:bs} according to
\begin{align}\label{eq:qft-definition-4d}
    |m\rangle_p\to \frac{1}{\sqrt{d}}\sum_{k\in\{b,c,d,e\}}\text{exp}\bigg[m\frac{2\pi i }{d}\Phi(p)\bigg]|m\rangle_k,
\end{align}
with $\Phi(p)$ denoting a path dependent phase, see supplementary. 
The FTI is a linear operation, but using linear optics only is not sufficient to project two photons into a $d$-dimensional Bell-state \cite{Calsamiglia2002}. Therefore, $d-2$ ancillary photons are added to the protocol. Finally, an extended unitary transformation in $U_{4+1}$ dimensions combined with post-selection of specific detection click-patterns is performed. To be noted that in $d$ dimensional Hilbert space, $d \times d$ outputs are required. Observing these click-patterns indicates the unambiguous projection of the two incoming photons into a four-dimensional Bell state. This completes the four-dimensional entanglement swapping protocol; further details are presented in the supplementary materials. To perform Bell measurements in dimension $d$, we here only use the $d$ heralding patterns that are known to provide perfect correlations after the entanglement swapping~\cite{luo2019quantum}. However, we note that many other heralding patterns are possible resulting in high-dimensional entanglement being shared between the parties, although not maximally entangled.  These could be used to significantly boost the success probability of the Bell measurement (see supplementary materials). 
\begin{figure*}[h!]
\centering
\includegraphics[width=1\textwidth]{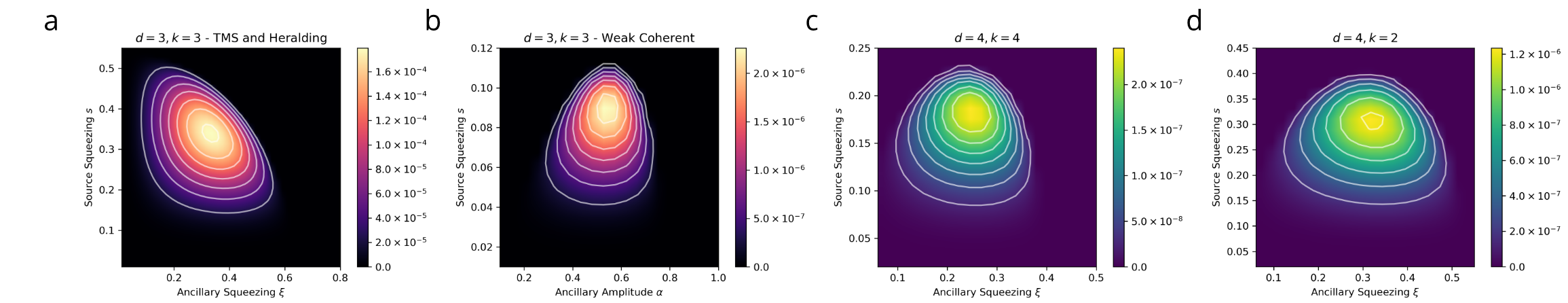}
\caption{{\bf Optimization of source and ancillary states parameters in the high-dimensional network.} Color-coded are the secrets bits per round of the simultaneous subspace coding protocol in different configurations and for different values of sources squeezing ($s$) and ancillary parameters.  a) $d=3$ and $k=3$ using an heralded single photon from a TMS state, with squeezing $\xi$ as ancilla. b) $d=3$ and $k=3$ using a weak-coherent state, with amplitude $\alpha$, for the ancilla. c) $d=4$ - $k=4$ and d) $d=4$ - $k=2$ using a TMS state with squeezing $\xi$ for the ancillas.}
\label{fig:skr_m}
\end{figure*}

\paragraph*{Practical ancillary quantum states}
Since the deterministic generation of single-photons in high-dimensions at 1550~nm is still an open challenge, we have proposed the possibility of replacing the ancillary single photons with ancillary quantum states which are easier to generate. As reported in Figure~\ref{fig:bs} we consider three different kind of quantum states for the ancillas: weak coherent states (WCS) two-modes squeezed states (TMS), and heralded single-photon source (HSPS). 

In particular, in the case of 3-dimensional systems, we propose to use weak coherent states generated from an attenuated laser. If the photon number per pulse is lower than one, on average we can approximate the weak coherent states with a single photon and successfully reveal the Hi-D swapping scheme.

Weak-coherent states cannot, however, be applied to a four-dimensional quantum states. In fact, because in the 4-dimensional HiD Bell measurement the number of additional particles needed are two, if two separate WCS are used the probability that two ancillary photons are generated is less than the probability of having two photons in just a single ancillary mode. The multi-photon contamination becomes thus significant in this scenario and reduces the measurement fidelity. However, we can exploit the photon-number correlations in two-mode squeezed states to mitigate such multi-photon noises between the two ancillary inputs. In fact, in the low squeezing regime, a TMS state approximate the ideal pair of ancillary photons with no intra-mode multiphoton contamination. Furthermore, the generation of a weakly squeezed quantum state is practical and experimentally simple to realize both using bulk and integrated optics.

Furthermore, for both $d=3$ and $d=4$, another possibility is to use heralded single photons from a TMS source as ancillas~\cite{kaneda2019high}. Note that HSPS can be also used with higher dimensionalities, but require twice the number of photons compared to the WC or TMS configurations considered above. 
%%%%%%%%%%%%%%
%maybe to be included in the discussion
%Note, the natural extension of this method to a larger dimensionality is not straightforward since the higher the dimensionality, the larger the number of required resources and the less efficient is the swapping operation.

%In fact, it is not possible to replace the $N-2$ single-photons with weak coherent states, since the probability of having in all of them a single-photon scales exponentially with the number of modes. The same principle can be applied to the weak squeezed states, since the probability of generating an $N-2$ squeezed state is very low.

We now have introduced the main building block of a single repeater quantum communication protocol based on high-dimensional quantum states and practical ancillary resources. We will now investigate the performance of the different systems as a function of the dimensionality of the Hilbert space.

\section*{Performance simulation}
% In this section we simulate how the scheme for high-dimensional Bell measurement described above performs when implemented in a realistic high-dimensional quantum network.

\paragraph*{Efficient scheme simulator}
In order to simulate the performance of a practical high-dimensional quantum network, we simulate the linear-optical circuit described in Fig.\ref{fig:bs} including the cases where two-mode squeezing, weak-coherent, or heralded single-photon states are used for the ancillas, and threshold detectors (non-number-resolving) are used at the outputs. A first challenge when building such simulator is that, when increasing the dimensionality $d$, the total number of photons increases significantly. This makes the calculation of the output detection probabilities rapidly inefficient for classical computers as the system size is increased. To maintain a good efficiency, we exploited the fact that, prior to the detection, the total initial state and its evolution can be fully described as Gaussian states and Gaussian transformations, which can be described efficiently~\cite{Weedbrook2012}. The non-Gaussian measurement with threshold detectors can then be simulated adapting techniques from Gaussian Boson Sampling~\cite{Hamilton2017}. In particular, to include weak-coherent states as possible inputs, we adopt a generalization of the Torontonian function used in Gaussian Boson Sampling to calculate output threshold detection probabilities for squeezed states~\cite{Quesada2018}, for the case where non-zero displacement is present~\cite{Bulmer2021}. Using this approach we are able to simulate qudit Bell measurement circuits with up to $>20$ detector clicks on a standard laptop, which we estimate to be compatible with high-dimensional quantum networks a dimensionality exceeding $d=15$ . This was fast enough to perform a detailed analysis of the circuits studied here, where we focus on cases with $d\leq 4$.

\begin{figure}[h]
\centering
\includegraphics[width=0.5\textwidth]{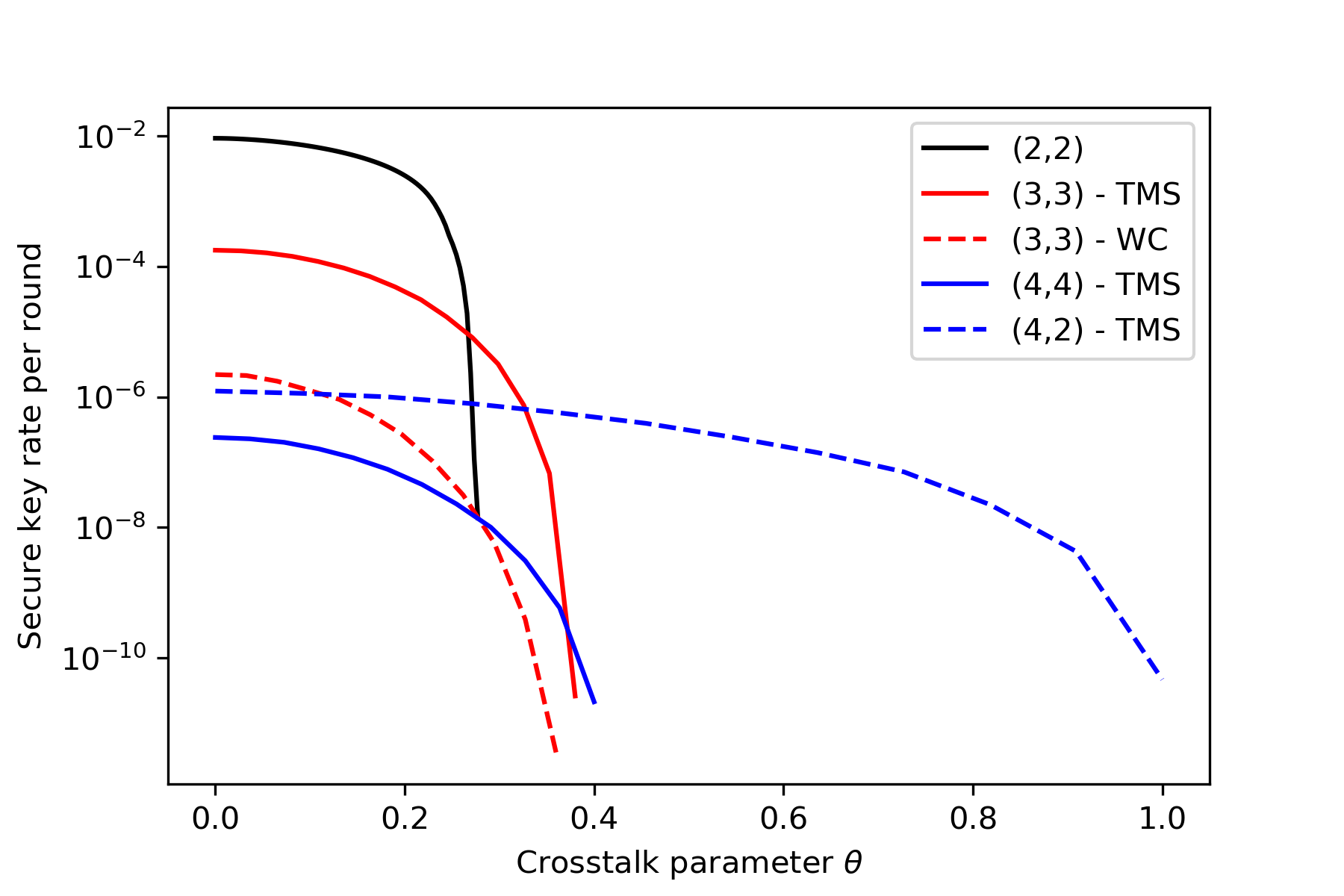}
\caption{{\bf Optimal secrets key rates in noisy quantum networks.} Represented are the secure bits per round against using the simultaneous subspace coding protocol as a function of the crosstalk noise parameter $\theta$. For each value of $\theta$, the key rates are optimized over all source and ancillary squeezing and/or amplitude parameters.}
\label{fig:RateVsParams_Marcus}
\end{figure}

\paragraph*{Entanglement based quantum key distribution through the network}
One of the most basic quantum communication protocols is quantum key distribution (QKD), in which two users generate a shared string of private random numbers. Using entanglement for this task unlocks the potential for device independent security and makes the system impervious to attacks on the source. Just as for qubits, the high-dimensional protocol for QKD works on the same principle: random local measurements are performed in multiple rounds on a shared entangled state. The rounds are divided into key rounds (in a predefined computational basis with high correlations) and test rounds (in multiple incompatible bases) for estimating privacy and the need for further classical error correction. While entanglement in high-dimensions is highly robust to noise, error correction in noisy states could still obliterate the actual key rates. Here, a recent protocol \cite{doda2020quantum} provides a workaround: simultaneously using multiple low-dimensional subspaces keeps the advantages of noise robust entanglement, while limiting error correction to lower-dimensional subspaces.
%\textcolor{blue}{Davide and Marcus: describe QKD protocol.}
In our work, we used the simultaneously coding technique for estimating the secret key rate between Alice and Bob for different Hilbert space dimensionality. 
In particular, we evaluate  the performance of our system as a function of the crosstalk parameter $\theta$, reported in Figure \ref{fig:RateVsParams_Marcus}. The secret key rate per round, in the asynthotic regime, has been calculated according to the following equation:

\begin{equation}
  K \geq H (X \lvert E_T) - H (X \lvert Y)   
\end{equation}
where $H (X \lvert E_T)$ is the von Neumann entropy of Alice's key round outcome X conditioned on the total information available to the eavesdropper Eve at the end of the parameter estimation procedure and and $H (X \lvert Y )$ is the conditional Shannon entropy between Alice’s and Bob’s key round outcomes\cite{doda2020quantum}.

\paragraph*{Optimising key rates for high-dimensional network.}
% \textcolor{blue}{Describe trade-offs between pumping and fidelities.}
In our proposal, the photons encoding the high-dimensional entangled states as well as the ancillary ones are generated probabilistically via TMS or coherent states. We define $s$ the two-mode squeezing parameters at the Alice and Bob entanglement sources, and $\xi$ ($\alpha$) the squeezing (displacement) parameters used for the ancillary photons. All these parameters affect the total key rate obtained from Alice and Bob after the high-dimensional entanglement swapping in two contrasting ways. On one hand, larger values indicate higher probabilities of generating photons thus initially increasing the rate. On the other hand, if the squeezing or displacement parameters are too large, multi-photon contamination becomes significant, amplifying noises and decreasing the effective fidelity of the shared state after entanglement swapping, reducing the secure key rate.
A central feature in the scheme is thus finding the optimal trade-off between these two processes, finding the parameters for Alice's and Bob's sources and ancillary photons that optimize the total key rate.

In Fig.~\ref{fig:skr_m} we numerically investigate the optimization for the networks in dimension $d=3$ and $d=4$. The key rate is obtained from the simulator for different values of the parameters, types of ancillary states, and encoding subspaces. It can be observed that the sources and ancillas parameters affect the rate in different ways, which also depends on the dimensionality and types of ancillas and encodings. The optimization is thus non-trivial for each configuration. Interestingly, configurations which are more robust to noises, such as the $k=2$ dimensional subspace encoding for $d=4$~\cite{doda2020quantum}, allows us to use higher squeezing parameters before multi-photon noises become significant, which significantly improves the optimal key rates.  

\paragraph*{Noise robustness}

One of the main features of high-dimensional QKD systems is an improved noise robustness compared to qubit-based systems~\cite{cozzolino2019review}. In order to identify regimes where the high-dimensional entanglement swapping scheme outperforms simpler qubit networks, it is import to analyse how performances are affected by realistic noises of practical relevance. 
%We numerically investigate this by adding three noise sources in our simulator: unitary linear-optical errors, losses, and dark-counts errors. 
We numerically investigate this by simulating noises coming from unitary linear-optical errors.
In particular, we consider a cross-talk model which can arise for example due to inter-mode contamination when transmitting OAM- and path-encoded qudits through fibers~\cite{Rademacher17, cozzolino2019orbital}. In this model, Alice's and Bob's idler photons (see Fig.~\ref{fig:bs}b) undergo an additional unitary evolution $\hat{U}=\exp(-i \hat{H} \theta)$, where $\hat{H} = \sum_{i=0}^{d-1} \ket{i}\bra{i+1 \mod d} + h.c.$ is a nearest-neighbour coupling Hamiltonian and the crosstalk parameter $\theta\in[0,\infty)$ embeds both the cross-talk coupling strength and the coupling length.

In presence of noise, we observe a change in the optimal sources parameters (Fig.~\ref{fig:skr_m}) which change for the noise level, as described in. Therefore, for a given characterised level of noise, new optimal source parameters have to be calculated. 
In Fig.~\ref{fig:RateVsParams_Marcus} we show the secure key rates per round, optimized over the source parameters, for different levels of crosstalk and dimensions. For low levels of noise the qubit-based entanglement swapping provides better rates compared to high-dimensional systems, due to the higher Bell-measurement success probability. However, using high-dimensional systems becomes advantageous due to their higher noise robustness. In fact, increasing the dimensionality allows us to achieve secure keys even in regimes where qubit-based schemes are no longer secure, although at lower rates. In particular, the $d=4$ and $k=2$ configuration shows a good rates even for very high levels of noises, promising for near-term networks where noises are expected to be significant.

\section*{Conclusions}
HiD quantum states in general exhibit special properties especially useful in practical quantum technologies experiencing noisy environments. In this article, we investigated the possibility of creating a repeater-based quantum network with HiD states using state-of-the-art devices. We have studied the expected performance of HiD systems and compared them to the standard qubit approach. Furthermore, we have presented a new tool for processing high-dimensional quantum states, which is useful for low dimensional spaces, i.e., $d<5$ in which ad-hoc quantum states can replace real single-photon sources for simplicity.

Our work represents the first concrete study of a possible quantum network based on high-dimensional quantum repeaters, but open questions remain. For example, we expect that the success probability of the HiD Bell state measurements at the repeater stage, an important factor reducing HiD key rates, can be significantly improved considering additional heralding patterns resulting in high-dimensional entangled states. Besides, computer aided design methods could be used to find more efficient schemes for HiD Bell state measurements~\cite{krenn2020computer}.

Furthermore, in our analysis we focused on the robustness to operational noise in high-dimensional networks compared to the simple qubit case, but not considered loss correction. We point out that losses affect the key rates as in the qubit case, and that quantum memories will thus be required in order to overcome the rate-distance limit~\cite{Pirandola2019}. The reduced success probability of Bell measurements in the HiD quantum repeaters means that longer coherence times for memories will be required. The improved noise robustness we showed here could compensate additional errors due to longer storage time. Further investigation on this point will thus require advances in HiD quantum memories, currently in their infancy~\cite{parigi2015storage, ding2016high, li2020high,tiranov2017quantification}. 

Our results promise an advantage in using HiD quantum repeaters for QKD networks in noisy environments. The results could open a pathway for practical applications where environmental noise cannot be neglected. Clearly, more work in this direction is needed to find a realistic trade-off for repeater-based QKD networks between dimensionality, secure key rates, loss- and noise tolerance.

\section*{Acknowledgements}
\vspace{-0.25cm}
The authors would like to thanks J. A. Adcock, B. Da Lio, D. Cozzolino, M. Krenn and M. Malik for the fruitful discussion.

\section*{Funding}
\vspace{-0.25cm}
This work is supported by the Center of Excellence, SPOC-Silicon Photonics for Optical Communications (ref DNRF123) and Hy-Q Center for Hybrid Quantum Networks (DNRF 139), by the EraNET Cofund Initiatives QuantERA within the European Union’s Horizon 2020 research and innovation program grant agreement No.
731473 (project SQUARE). M.H. acknowledges funding from the Austrian Science Fund (FWF)
through the START project Y879-N27. M.E. acknowledges support from FWF project W 1210-N25 (CoQuS).

%\newpage
%\bibliography{mybib}{}
%\bibliographystyle{plain}
%\bibliographystyle{unsrt}
\bibliographystyle{ieeetr}

\end{document}